\documentclass[12pt,aps,prb,preprint]{revtex4}   % style for Physical
\usepackage{amsmath}    % need for subequations
\usepackage{amsfonts}  %note how statements can be commented out
\usepackage{amssymb}
\usepackage{graphicx}   % for figures
\usepackage{color}
\newcommand{\be}{\begin{equation}}
\newcommand{\ee}{\end{equation}}
\newcommand{\bea}{\begin{eqnarray}}
\newcommand{\eea}{\end{eqnarray}}
\newcommand{\nn}{\nonumber}

\newcommand{\gp}{\left( \gamma\cdot \Pi \right)}

\begin{document}

\title{The Electron Propagator in External Electromagnetic Fields in
  Lower Dimensions}
\author{Gabriela Murgu\'{\i}a} 
\affiliation{Departamento de F\'{\i}sica, Facultad de Ciencias, Universidad
  Nacional Aut\'onoma de M\'exico. Apartado Postal 21-092, 
  M\'exico, D.F., 04021, M\'exico.}
 \email{murguia@ciencias.unam.mx}   %optional

\author{Alfredo Raya}
\affiliation{Instituto de F\'{\i}sica y Matem\'aticas, Universidad
  Michoacana de San Nicol\'as de Hidalgo. Apartado Postal 2-82,
  Morelia, Michoac\'an, 58040, M\'exico.}
\email{raya@ifm.umich.mx,ansac@ifm.umich.mx}

\author{Edward Reyes}
\affiliation{Instituto de Ciencias Nucleares, Universidad Nacional
  Aut\'onoma de M\'exico.  Circuito Exterior s/n, Ciudad
  Universitaria, M\'exico, D.F., 04510, M\'exico.}
\altaffiliation[Also at ]{Instituto de F\'{\i}sica y Matem\'aticas, 
  Universidad Michoacana de San Nicol\'as de Hidalgo. Apartado Postal 2-82,
   Morelia, Michoac\'an, 58040, M\'exico.}  %  optional
\email{edward.reyes@nucleares.unam.mx}

\author{\'Angel S\'anchez}
\affiliation{Instituto de F\'{\i}sica y Matem\'aticas, Universidad
  Michoacana de San Nicol\'as de Hidalgo. Apartado Postal 2-82,
  Morelia, Michoac\'an, 58040, M\'exico.}
\email{ansac@ifm.umich.mx}

\date{\today}

\begin{abstract}
We study the electron propagator in quantum electrodynamics in lower
dimensions. In the case of free electrons, it is well known that the
propagator in momentum space takes the simple form
$S_F(p)=1/(\gamma\cdot p-m)$. In the presence of external
electromagnetic fields, electron asymptotic states are no longer
plane-waves, and hence the propagator in the basis of momentum
eigenstates has a more intricate form. Nevertheless, in the basis of
the eigenfunctions of the operator $\gp^2$, where $\Pi_\mu$ is the
canonical momentum operator, it acquires the free form
$S_F(p)=1/(\gamma\cdot \overline{p}-m)$ where $\overline{p}_\mu$ depends on
the dynamical quantum numbers. We construct the electron propagator in
the basis of the $\gp^2$ eigenfunctions. In the (2+1)-dimensional
case, we obtain it in an irreducible representation of the Clifford
algebra incorporating to all orders the effects of a magnetic field of
arbitrary spatial shape pointing perpendicularly to the plane of
motion of the electrons. Such an exercise is of relevance in graphene
in the massless limit. The specific examples considered include the
uniform magnetic field and the exponentially damped static magnetic field. We
further consider the electron propagator for the massive Schwinger
model incorporating the effects of a constant electric field to all
orders within this framework.
\end{abstract}

\maketitle

\section{Introduction}

A powerful technique to solve wave equations in quantum theory is the
Green's function method. The evolution of the wave functions is
encoded in the propagators, which in plain words are the inverses of
the differential wave operators. For free particles, which are
described by plane-waves, these propagators acquire simple diagonal
forms in momentum space due to the isotropy of the space.  However,
when the isotropy is lost, particles cease to be described as
plane-waves and their corresponding propagators have more intricate
forms in the basis of momentum eigenstates, sometimes at the point of
being intractable for actual calculations. An example of such systems
is a gas of electrons subject to an intense magnetic field in a fixed
direction. The non-relativistic quantum problem for this system, known
as the Landau problem,~\cite{Landau} reveals interesting features of
the dynamics of electrons in external magnetic fields.  On one hand,
the Lorentz force lacks of component along the direction of the field. This makes a clear separation between the parallel (free) and
transverse (dynamical) components of the trajectories of the electrons with respect to the field lines. On the
other hand, in the transverse plane, the trajectories are confined
around the field lines. In such a case, the energy levels, called
Landau levels, develop a discrete spectrum. These features combined translate into a 
rather intricate form of the electron propagator~\cite{brasileiros} as compared with its free
counterpart. For relativistic electrons, the problem does not get any
simpler. Yet,  the task of unveiling these propagators
for systems just as complicated  is  worthwhile.

There have been a number of strategies developed in the past to derive
the propagator for relativistic electrons in background
electromagnetic fields, perhaps the best known being the
Fock-Schwinger ``proper time''
  method,~\cite{Fock,Schwinger} which was originally developed in
relativistic quantum field theory (see Ref.~[5] for a brief discussion of the method), but that has been applied to a
number of problems in ordinary non-relativistic quantum
mechanics.~\cite{brasileiros,otros} However, there have been
interesting alternatives to the Fock-Schwinger method to
incorporate the effects of the background fields into the propagators,
like the second quantization of solutions to the Dirac equation in the
background fields,~\cite{Kaushik} path integral methods,~\cite{path}
and the Ritus eigenfunction method.~\cite{Ritus} In this article, we
derive the massive electron propagator in background electromagnetic
fields in lower space-time dimensions through the later method, which
is based upon the diagonalization of the Dirac operator on the basis
of the eigenfunctions of the operator $\gp^2$, with $\Pi_\mu$ being the
canonical momentum operator.

We begin this article by considering the $d$-dimensional free electron
propagator in Sect.~\ref{Free_Dirac_Propagator}. The details of the
Ritus method are presented in
Sect.~\ref{Propagator_Magnetic_Fields}. For the sake of simplicity, we
present the study of the electron propagator in
(2+1)-dimensional QED, where a third spatial dimension is
suppressed. This is not a mere theoretical simplification, and we
explain ourselves: back in time, some twenty years
ago,~\cite{Semenoff} it was shown that the low-energy effective theory
of graphene in a tight-binding approach is the theory of two species
of massless Dirac electrons in a (2+1)-dimensional Minkowski
space-time, each on a different irreducible representation of the
Clifford algebra. The isolation of graphene samples~\cite{Novoselov}
in 2004 and 2005, has given rise to the new paradigm of relativistic
condensed matter,~\cite{graphene} bringing a new boost, both
theoretical and experimental, to the matching of common interests of
the condensed matter and high energy physics communities. Thus, the
massless limit of our findings is of direct relevance in this
subject.~\cite{planarqed} We assume the electrons moving in a magnetic
field alone pointing perpendicularly to their plane of motion. We
first develop the general case and then, in Sect.~\ref{Examples} we
present a couple of examples: the motion of electrons in a uniform
magnetic field, which is a canonical example to present the Ritus
method,~\cite{Khalilov} and the case of a static
magnetic field which decays exponentially along the $x$-axis. The effects of such a magnetic field to all orders can hardly be
considered with other approaches, but it can be straightforwardly
studied within the Ritus method.

Electric field configurations can also be studied in
this framework. With this goal in mind, in
Sect.~\ref{Propagator_Electric_Fields} we analyze the electron
propagator in a constant electric field in the massive Schwinger
model~\cite{Schwinger}, that is, quantum electrodynamics in
(1+1)-dimensions. The Schwinger model exhibits spontaneous symmetry
breaking of the $U(1)$ symmetry, as well as confinement
of electrons, thus is a favorite toy model to study some aspects of
quantum chromodynamics. These properties can be studied from the
electron propagator, which acquires a free form when expressed in the
basis of the eigenfunctions of the operator $\gp^2$. 
We settle the basis for such studies in this article. In Sect. VI
we present a discussion and final remarks. At the end, we
propose a project for the interested reader, in which the Ritus method has to be
applied in order to obtain the electron propagator in a different situation than the considered here.

\section{Free Dirac Propagator}
\label{Free_Dirac_Propagator}

We start our discussion from the equation of motion for free
electrons, namely, the Dirac equation~\cite{fn1}
\be
(i\gamma\cdot \partial -m)\psi(x)=0,\label{freeDirac}
\ee
where the $\gamma^\mu$-matrices fulfill the Clifford algebra $\{
\gamma^\mu,\gamma^\nu\}=2g^{\mu\nu}$, where
$g^{\mu\nu}=\text{diag}(1,-1,\ldots,-1)$, depending upon the number $d$ of
space-time dimensions considered.

In order to solve the Dirac equation, we introduce the Green's function or propagator
for this wave equation.
Propagators are important because they allow to have a visual description of interaction processes out of which we desire  
to calculate several dynamical quantities, like decay rates, 
scattering cross sections, etcetera. 
In scattering problems, for instance, 
%the attention is focused upon wave solutions 
%which develop in time from initial conditions imposed in the remote 
%past rather than on stationary energy eigenfunctions, that is, standing 
%waves. Characteristically, given 
we consider a wave packet which in the remote 
past represents a particle approaching a potential, and we wonder how the 
wave will look like in the remote future, after the interaction with the potential. For the sake of illustration, let us visualize
this process in ordinary quantum mechanics.
%One easily way to understand this process is by mean of Huygens' principle. 
Imagine that we know the wave function $\psi(x,t)$ at one particular 
time $t$. % It may be found at any later time $t'$ by considering 
%at time $t$ 
At this moment, each point of space $x$ may be viewed  as a source of spherical waves propagating outwards 
%from $x$. 
in such a way that the strength of the wave amplitude arriving at point $x'$ at 
a later time $t'$ 
%from the point $x$ 
will be proportional to the original wave 
amplitude $\psi(x,t)$. This is Huygens' principle at work. If we denote the constant of proportionality by 
$iG(x',t';x,t)$, the total wave arriving at the point $x'$ at time $t'$ will be
\begin{equation}
    \psi(x',t')=i\int d^3x \ G(x',t';x,t)\psi(x,t)\;.
\end{equation}
Here, $G(x',t';x,t)$ is precisely the Green's function or propagator. 
Thus, the knowledge of the Green's function enables us to construct the physical 
state which develops in time from any given initial state. In a theory 
without interactions the propagator is referred to as the free propagator, $G_{free}(x',t';x,t)$.
However if a potential $V(x_1,t_1)$  is turned on at a time $t_1$ at the point $x_1$ for an interval $\Delta t_1$, then the wave 
function and propagator will be modified as 
\begin{eqnarray} 
     \psi(x',t')&=&\psi_{free}(x',t')+\int d^3x_1 \ G_{free}(x',t';x_1,t_1)V(x_1,t_1)\psi_{free}(x_1,t_1) \Delta t_1\;,
                 \nonumber \\
     G(x',t';x,t)&=&G_{free}(x',t';x,t)\nonumber\\
&&+ \int d^3x_1 \ G_{free}(x',t';x_1,t_1)V(x_1,t_1)G_{free}(x_1,t_1;x,t) \Delta t_1\,,
\end{eqnarray}
where the first term in the last expression represents the propagation of a free particle from $(x,t)$ to 
$(x',t')$ and the second term, read from right to left, represents free propagation from 
$(x,t)$ to $(x_1,t_1)$, followed by a scattering at $(x_1,t_1)$, and free propagation from $(x_1,t_1)$ 
to $(x',t')$. The same idea might be straightforwardly translated to the relativistic case.

In order to find the Green's function $G(x,x')$ for the Dirac
equation~(\ref{freeDirac}), we consider
\be
(i\gamma\cdot\partial-m)G(x,x')=\delta(x-x').
\ee
Since the Green's function is translationally invariant,
$G(x,x')=G(x-x')$, we can find its representation in momentum space
through its Fourier transformation, $S_F(p)$, defined as
\be
G(x,x')=\int \frac{d^dp}{(2\pi)^d} e^{-ip\cdot x} S_F(p) (e^{-ip\cdot x'})^*,
\ee
the asterisk denoting complex conjugation. 
%\begin{quote}
%{\bf Excersice 1.} 
Upon acting with the Dirac operator on $G(x,x')$, we have that
%and making use of the representation of the $d-$dimensional $\delta$-function in terms of plane-waves,
\bea
(i\gamma\cdot\partial-m)G(x,x') &=&
(i\gamma\cdot\partial-m) \int \frac{d^dp}{(2\pi)^d} e^{-ip\cdot x} S_F(p)(e^{-ip\cdot x'})^*\nn\\
&=&
\int \frac{d^dp}{(2\pi)^d} e^{-ip\cdot x} (\gamma\cdot p-m)S_F(p) (e^{-ip\cdot x'})^*\nn\\
&=&   \int \frac{d^dp}{(2\pi)^d} e^{-ip\cdot x}(e^{-ip\cdot x'})^* .
\eea
Here, in the last line we have made use of the representation of the $d-$dimensional $\delta$-function in terms of plane-waves. Hence, we
see that
\be
S_F(p)=\frac{1}{\gamma\cdot p - m}\;.\label{freeprop}
\ee
%\end{quote}
This is the free electron propagator in momentum space, suitable for
calculations in quantum electrodynamics.

In the next section we consider the motion of electrons in magnetic
fields. A method shall be presented to render the propagator similar
in form to the free propagator given above.

\section{Propagator in Magnetic Fields}
\label{Propagator_Magnetic_Fields}

The discussion in this section takes place in a (2+1)-dimensional
Minkowski space-time with metric $g_{\mu\nu}=g^{\mu\nu}=\text{diag}(1,-1,-1)$
with a system of natural units $\hbar=\tilde{c}=1$ where $\tilde{c}$
is the planar ``speed of light'' (Fermi velocity), which is some two
orders of magnitude smaller than $c$.~\cite{graphene,vF} As we
mentioned before, this is not a mere theoretical simplification. On
the contrary, we aim to describe the low-energy theory of graphene in
the massless limit. We start from the free Dirac equation shown in
Eq.~(\ref{freeDirac}), for which we only need three $\gamma^\mu$-matrices.  The lowest dimensional
representation of these matrices is $2\times 2$, and hence we can
choose them to be proportional to the Pauli matrices,
%\begin{quote}
%{\bf Excercise 2.} Verify that the set of matrices
\be  \gamma^{0} = \sigma_3 , \hspace{10mm} \gamma^{1}  =
i\sigma_1 , \hspace{10mm} \gamma^{2}  = i\sigma_2     \; ,
\label{primera}
\ee
which verify
$\gamma^\mu\gamma^\nu=g^{\mu\nu}-i\epsilon^{\mu\nu\lambda}\gamma_\lambda$,
with $\gamma_\lambda=g_{\lambda\mu}\gamma^\mu$.
%\end{quote}
Notice that there exists a second inequivalent representation of the
Clifford algebra, specified by the $\tilde{\gamma}^\mu$-matrices
\be
\tilde{\gamma}^{0} = \sigma_3 ,
\hspace{10mm} \tilde{\gamma}^{1}  = i\sigma_1 , \hspace{10mm} \tilde{\gamma}^{2}  =
-i\sigma_2     \; . \label{segunda}
\ee
%This
%representation merely reverses the sign of one of the spatial
%$\gamma^i$ matrices, and thus the solutions to the momentum-space
%Dirac equation can be derived from those in
%eq.~(\ref{freeDirmomentum}) through the replacement
%$u_p \to \sigma_1 u_p$, and represent a particle solution with spin down, and
%antiparticle solution with spin up.
In graphene, the two representations describe two different
electron species in each of the two triangular
sub-lattices of the honeycomb lattice. These are conveniently merged
in a 4-component spinor with a reducible representation of the Dirac
matrices given, for example, as \be \gamma^{0} =
\left( \begin{array}{cc}\sigma_3 & {\phantom{-}} 0 \\ 0 &
  -\sigma_3 \end{array}\right)\;,
\hspace{10mm} \gamma^i  = i \left( \begin{array}{cc}\sigma_i  & {\phantom{-}} 0 \\ 0 & -\sigma_i \end{array}\right)\;,\label{red}
\ee
$i=1,2$. For the purposes of this article, we shall only consider the
representation given in Eq.~(\ref{primera}), but the generalization is
straightforward, and will be left as an exercise for the interested reader.

In a background electromagnetic field, the Dirac equation takes the
form
\be
\left(\gamma\cdot\Pi-m\right)\Psi=0, \label{fieldDirac}
\ee
where $\Pi_\mu = i\partial_\mu+eA_\mu$ and $A_\mu$ is the
electromagnetic potential defining the external field.  Let us
consider a magnetic field alone pointing perpendicularly to the plane
of motion of the electrons, in such a fashion that, working in a
Landau-like gauge, $A^\mu=(0,0,W(x))$, where $W(x)$ is some function such
that $W'(x)=\partial_xW(x)$ defines the profile of the field.
%Feynman and Gell-Mann~\cite{Feynman} realized that the 4-component
%Dirac equation in a background electromagnetic field can be solved
%by a wave function written in  terms of 2-component spinors as \be
%{\color{red} \Psi=esto solo esta hecho para el caso de campo
%constante, aun no para un campo earbitrario...} \ee

As before, we are interested in solving Eq.~(\ref{fieldDirac}) through the Green's
function method. In the presence of an external electromagnetic  field, the
Green's function $G(x,x')$ satisfies   
\begin{equation}
 \left((\gamma\cdot\Pi)-m\right)G(x,x')=\delta(x-x').
\label{method1}
\end{equation}
Since $(\gamma\cdot\Pi)$ does not commute with the momentum
operator, neither the wave function nor $G(x,x')$ can be
expanded in plane-waves, and this does not allow to have a diagonal
propagator in momentum space. The scheme we choose to deal with the
external fields was developed by Ritus.\cite{Ritus} 
The crucial point is to realize that the Green's function above should be a combination of all scalar structures obtained by contracting  the $\gamma^\mu$-matrices,  the canonical momentum
$\Pi_\mu$ and the electromagnetic field strength tensor $F_{\mu\nu}=[\Pi_\mu,\Pi_\nu]/e\equiv\partial_\mu A_\nu - \partial_\nu A_\mu$, which are  compatible with Lorentz symmetry, gauge invariance and charge conjugation properties of quantum electrodynamics. In our conventions, we have that 
\begin{equation}
    G(x,x')=G(\gamma_\mu\Pi^\mu,\sigma^{\mu\nu}F_{\mu\nu},
             (\tilde{F}^{\nu}\Pi_\nu)^2)
\label{method2}
\end{equation}
where $\sigma_{\mu\nu}=i[\gamma_\mu,\gamma_\nu]/2$ and
\begin{eqnarray}
      \tilde{F}_{\mu}\equiv 
            \frac{1}{2}\epsilon_{\mu\nu\alpha}F^{\nu\alpha} 
\label{mathod3}
\end{eqnarray}
is the dual field strength tensor, which in (2+1)-dimensions is simply a vector. The key observation is that all the above structures commute with
$(\gamma\cdot\Pi)^2$, that is
\begin{eqnarray}
 [(\gamma\cdot\Pi)^2,(\gamma\cdot\Pi)]=
 [(\gamma\cdot\Pi)^2,\sigma^{\mu\nu}F_{\mu\nu}]=  
 [(\gamma\cdot\Pi)^2,(\tilde{F}^{\mu}\Pi_\mu)^2]=0\;,
 \label{method4}
\end{eqnarray}
relations which are straightforward to prove and represent an excellent exercise we invite the reader to carry out. As a consequence of the above relations, we have that
\begin{equation}
     \left[(\gamma\cdot\Pi)^2,G(x,x')\right]=0\;.
\label{method5}
\end{equation}
This is the main point in the Ritus method, since we know
that commutating operators have simultaneous eigenfunctions. For our particular problem,  this statement allows us to
expand the Green's function $G(x,x')$ in the same basis of eigenfunctions of $(\gamma\cdot\Pi)^2$. Furthermore,  if we perform a similarity transformation on
$(\gamma\cdot\Pi)^2$ in which  it acquires a diagonal form in momentum space, then
the same transformation makes the Green's function diagonal too. This is a
general statement with operators, and hence the Ritus method can be
applied to derive the propagator of charged scalar~\cite{scalar} and vector~\cite{vector} particles in a similar way.

The similarity transformation that makes $(\gamma\cdot\Pi)^2$ diagonal
in the momentum space is such that 
\begin{equation}
 \mathbb{E}_p^{-1}(\gamma\cdot\Pi)^2 \mathbb{E}_p = p^2 \mathbb{I}\;,\label{simtransf}
\end{equation}
where $\mathbb{E}_p$ are the transformation matrices, $\mathbb{I}$ is the unit matrix and $p^2$ can be any real number. Therefore, when we apply 
$\mathbb{E}_p$ functions to the propagator, it will become diagonal in momentum
space.  

It is important to notice that in the fermionic case, the spin operator is realized in terms of the 
$\gamma^\mu$-matrices, and thus the $\mathbb{E}_p$ functions inherit its matrix form. For different charged particles, the spin operator is realized in a different ways. For example, for scalar particles, the $\mathbb{E}_p$ functions are simply scalars,~\cite{scalar} whereas in the case of charged gauge bosons, the spin
structure is embedded in a Lorentz tensor, and therefore the
$\mathbb{E}_p$ functions also comply a Lorentz tensor structure.\cite{vector}

Our task now is to study the structure of the $\mathbb{E}_p$ matrices for the case of Dirac fermions in (2+1)-dimensions. The similarity transformation~(\ref{simtransf}) can be written  as 
\begin{equation}
(\gamma\cdot\Pi)^2\mathbb{E}_p = p^2 \mathbb{E}_p\;,
\label{eigen}
\end{equation}
which is an eigenvalue equation for the matrices $\mathbb{E}_p$, which we shall refer to as the Ritus eigenfunctions.
Explicitly, we have that 
\bea
\gp^2 &=& \gamma^\mu\gamma^\nu\Pi_\mu \Pi_\nu \nn\\
 &=&\left(\frac{1}{2} \left[\gamma^\mu,\gamma^\nu \right]+\frac{1}{2}\left\{\gamma^\mu,\gamma^\nu\right\}\right)
\left(\frac{1}{2} \left[\Pi_\mu,\Pi_\nu \right]+\frac{1}{2}\left\{\Pi_\mu,\Pi_\nu\right\}\right)\nn\\
&=& \left( -i \sigma^{\mu\nu} +g^{\mu\nu}\right)\left(\frac{ie}{2}F_{\mu\nu}+\frac{1}{2}\left\{\Pi_\mu,\Pi_\nu\right\}\right)\nn\\
&=& \Pi^2+\frac{e}{2}\sigma^{\mu\nu}F_{\mu\nu}\;.
\eea
%where $\sigma^{\mu\nu}=i[\gamma^\mu,\gamma^\nu]/2$ and $F_{\mu\nu}=[\Pi_\mu,\Pi_\nu]/e\equiv \partial_\mu A_\nu - \partial_\nu A_\mu$ is the usual electromagnetic field strength tensor.
%\end{quote}
For the configuration of the field we are considering, the only
non-vanishing elements of the field strength tensor are
$F_{12}=-F_{21}=W'(x)$. Moreover, in the
representation~(\ref{primera}) of the Dirac matrices,
$\sigma^{12}=\sigma_3$, and hence the ${\mathbb E}_p$ functions
satisfy
\be
\left(\Pi^2+e\sigma_3W'(x)\right) {\mathbb E}_p = p^2 \ {\mathbb E}_p\;.
\ee
In order to solve this equation, the key observation comes from the
fact that the operator $(\gamma\cdot \Pi)$ satisfies
\be
\left[ \gp, i\partial_t\right]= \left[ \gp,-i\partial_y\right]= \left[ \gp, {\cal H}\right]=0\;,
\ee
with ${\cal H}=-\gp^2+\Pi_0^2$. Thus,  these operators share a complete
set of eigenfunctions, namely, ${\mathbb E}_p$. Let the eigenvalues of
these operators be
\be
i\partial_t {\mathbb E}_p= p_0 {\mathbb E}_p\;,\qquad
i\partial_y {\mathbb E}_p= -p_2 {\mathbb E}_p \;, \qquad
{\cal H} {\mathbb E}_p= k {\mathbb E}_p \;.\label{eigenvalues}
\ee
These eigenvalues label the solutions to the massless Dirac equation in the
background field. Furthermore, from Eq.~(\ref{eigen}), we have that
$p^2=p_0^2-k$. Hence, the ${\mathbb E}_p$ functions verify
\be
\left(-\Pi_1^2-\Pi_2^2+e\sigma_3 W'(x)\right){\mathbb E}_p= -k {\mathbb E}_p\;.\label{opeq}
\ee
The first two terms of the operator on the l.h.s. of this equation act
on the orbital degrees of freedom of the eigenfunctions ${\mathbb
  E}_p$, whereas the last term acts only in its spin degrees of
freedom. Hence we can make the ansatz
\be
{\mathbb E}_p = E_{p,\sigma}\  \omega_\sigma\;, \label{an1}
\ee
where $\omega_\sigma$ is the matrix of eigenvectors of $\sigma_3$ with eigenvalues
$\sigma=\pm 1$, respectively. Furthermore, since the above operator equation is independent of $t$ and $y$, we can look for solutions of the form
\be
E_{p,\sigma}=N_\sigma e^{-i(p_ot-p_2y)}F_{k,p_2,\sigma}\;, \label{an2}
\ee
with $N_\sigma$ being the corresponding normalization constant. Substituting the ans\"atze~(\ref{an1}) and~(\ref{an2}) into Eq.~(\ref{opeq}), we arrive at
\be
\left(\partial_x^2-(-p_2+eW(x))^2+e\sigma W'(x)\right) F_{k,p_2,\sigma}= -k F_{k,p_2,\sigma}\;,\label{pauli}
\ee
which is the Pauli Hamiltonian with the constrained vector
potential, mass $m=1/2$ and gyromagnetic factor $g=2$, and turns out
to be supersymmetric in the Quantum Mechanical sense (SUSY-QM).~\cite{susyqm}

From the solutions to the above equation, we can construct the Ritus
eigenfunctions ${\mathbb E}_p$ as
\be
{\mathbb E}_p=\left(\begin{array}{cc} E_{p,1}(z) & 0 \\ 0 & E_{p,-1}(z)\end{array} \right)\;,
\label{ritus}
\ee
where $p=(p_0,p_2,k)$ and $z=(t,x,y)$. 
%
%The $E_{p,\sigma}$ functions have the form
%\be
%E_{p,\sigma}=N_\sigma e^{-i(p_ot-p_2y)}F_{k,p_2,\sigma}\;,
%\ee
%where the $F_{k,p_2,\sigma}$ are the unnormalized solutions to
%eq.~(\ref{pauli}) and $N_\sigma$ the corresponding normalization
%constant.
%
Being a complete set, the eigenfunctions ${\mathbb E}_p$
given in Eq.~(\ref{ritus}),  satisfy
\bea
\int dz \ \overline{\mathbb E}_{p'}(z) {\mathbb E}_p(z)&=& {\mathbb I}\delta(p-p')\;,\nn\\
\int dp \ {\mathbb E}_{p}(z) \overline{\mathbb E}_p(z')&=& {\mathbb I} \delta(z-z')\;,
\eea
with $\overline{\mathbb E}_p(z)=\gamma^0 {\mathbb E}_p^*(z)\gamma^0$ and ${\mathbb I}$ is the $2\times 2$ unit matrix.
  
The three-momentum $\overline{p}_\mu=(p_0,0,\sqrt{k})$ plays an important role in the method. Its definition involves the dynamical quantum numbers $p_0$ and $k$, but not $p_2$, which merely fixes the origin of the $x$
coordinate.  This means that in the Ritus method the propagator is written only in terms of the eigenvalues of the dynamical operators commuting with $\gp$. Notice that this special three-vector verifies $\overline{p}^2=p_0^2-k=p^2$, and it is defined through the  relation
\be
\gp {\mathbb E}_p = {\mathbb E}_p (\gamma\cdot \overline p)\;.\label{p_barra}
\ee
Indeed, we can prove this relation since, in matrix form,   
\begin{eqnarray}
\gp {\mathbb E}_p &=&
\left(\begin{array}{cc} i\partial_t E_{p,1} & \lbrack -\partial_x-(-i\partial_y-eW)\rbrack E_{p,-1} \\ \lbrack -\partial_x + (-i\partial_y-eW)\rbrack E_{p,1} & -i\partial_t E_{p,-1}
\end{array}\right)  \nn\\
&=&
\left(\begin{array}{cc} p_0 E_{p,1} & \lbrack -\partial_x-(p_2-eW)\rbrack E_{p,-1} \\ \lbrack -\partial_x + (p_2-eW)\rbrack E_{p,1} & -p_0 E_{p,-1}
\end{array}\right)
\;,
\end{eqnarray}
 and 
\be
{\mathbb E}_p (\gamma\cdot \overline{p})=\left(\begin{array}{cc} p_0 E_{p,1} & -\sqrt{k} E_{p,1} \\ \sqrt{k}E_{p,-1} & -p_0 E_{p,-1}
\end{array}\right).
\ee
Here, we have made use of the properties of the ${\mathbb E}_p$ functions, Eq.~(\ref{eigenvalues}). We immediately observe that the relation holds for the diagonal components, while the off-diagonal elements lead to the system of equations 
\bea
\lbrack-\partial_x - (p_2-e W) \rbrack E_{p,-1} &=&-\sqrt{k} E_{p,1}\;, \label{asenso}\nn\\
\lbrack - \partial _x + (p_2-e W) \rbrack E_{p,1} &=& \sqrt{k} E_{p,-1}\;. \label{desenso}
\eea
If we act on the left of the second of the above expressions with the differential operator on the l.h.s. of the first one, we have
%si aplicamos el operador del lado izquierdo de la ecuacion (\ref{asenso}) a la ecuacion (\ref{desenso}), obtenemos
\bea
\left[-\partial_x-(p_2-eW)\right] \left[-\partial_x+(p_2-eW)\right]E_{p,1}&=&
\left[-\partial_x-(p_2-eW)\right]\sqrt{k} E_{p,-1} \nn \\
\left[ \partial_x^2 - (p_2 -eW)^2+ e\partial_x W -eW\partial_x \right] E_{p,1} &=& -k E_{p,1} \nn \\
\left[\partial_x^2- (p_2 -eW)^2 +eW'  \right]E_{p,1} &=& -kE_{p,1}
\eea
which is nothing but Eq.~(\ref{pauli}) for $E_{p,1}$. Analogously, acting on the left of Eq.~(\ref{asenso}) with the differential operator on the l.h.s. of Eq.~(\ref{desenso}), we obtain
%lo cual es la la ecuacion que satisface $\mathbb{E}_{p,1}$ (ecuaci'on \ref{ec_pauli}). Analogamente, si aplicamos el operador que esta del lado izquierdo en la ecuaci'on (\ref{desenso}) a la ecuaci'on (\ref{asenso}), obtenemos
\be
[\partial_x^2+ (p_2 -eW)^2 -eW'  ]E_{p,-1} = -kE_{p,-1},
\ee
which is again Eq.~(\ref{pauli}), now for $E_{p,-1}$. Thus the relation~(\ref{p_barra}) is also valid for the off-diagonal elements. 
%con lo cual queda demostrado la proposici'on (\ref{p_barra}).

%Below we shall obtain an expression for the fermion propagator in the
%background magnetic field in the ${\mathbb E}_p$ basis.

%\section{Ritus Propagator}

With the ${\mathbb E}_p$ functions, we can consider the Green's
function method to obtain the propagator in the presence of the
field. We start from Eq.~(\ref{method1}) and re-label $(x,x')\to(z,z')$. Then we define the Green's function in momentum space as
\be
G(z,z')=\int dp \ {\mathbb E}_p(z) S_F(p) \overline{\mathbb E}_p(z')\;.
\ee

We would like to stress that the integration might as well represent a
sum, depending upon the continuous or discrete nature of the components
of the momentum.
%\begin{quote}
%{\bf Excercise 4.} 
Applying the Dirac operator $\left(\gamma\cdot\Pi -m\right)$ to $G(z,z')$, we have that
%and making use of the properties of the ${\mathbb E}_p$ functions and representation of the $\delta$-function in this basis, show that
\bea
\left(\gamma\cdot \Pi-m\right )G(z,z')&=& (\gamma\cdot\Pi-m)\int dp \ {\mathbb E}_p(z) S_F(p) \overline{\mathbb E}_p(z')\nn\\
&=& \int dp \ (\gamma\cdot\Pi-m) {\mathbb E}_p(z) S_F(p) \overline{\mathbb E}_p(z')\nn\\
&=&\int dp \ {\mathbb E}_p(z)(\gamma\cdot\overline{p}-m)  S_F(p) \overline{\mathbb E}_p(z')\nn\\
&=&\int dp \ {\mathbb E}_p(z) \overline{\mathbb E}_p(z')\;,
\eea
where in the last line we have used the representation of the $\delta$-function in the ${\mathbb E}_p$ basis. Hence we notice that 
the Ritus propagator takes the form of a free propagator, namely,
\be
S_F(p)=\frac{1}{\gamma\cdot \overline{p}-m}\;,
\ee
in this basis, with $\overline{p}_\mu$ defined through Eq.~(\ref{p_barra}). 
%\end{quote}

On physical grounds, the ${\mathbb E}_p$ functions correspond to the
asymptotic states of electrons with momentum $\overline{p}$ in the background of the external field under consideration. With the 
help of these functions and the property~(\ref{p_barra}), we can find the solutions of
the Dirac equation~(\ref{fieldDirac}) in a straightforward manner. To this end, we propose $\Psi = {\mathbb E}_p
u_{\overline{p}}$, then 
\be
(\gamma \cdot \Pi - m) \mathbb {E}_p u_{\overline {p}}  = \mathbb {E}_p (\gamma \cdot \bar {p} - m)  u_{\overline {p}}= 0,
\ee
and thus we see that $u_{\overline{p}}$ is simply a free spinor describing an electron with momentum $\overline{p}$.
Notice that with this form of $\Psi$, the information concerning the
interaction with the background magnetic field has been factorized
into the ${\mathbb E}_p$ functions and throughout the $\overline{p}$ dependence of $u_{\overline{p}}$.
%
%Thus, given a magnetic field configuration, solving for the Ritus eigenfunctions of the operator $(\gamma \cdot \Pi)^2$, and for the free (massive) Dirac spinors, the full solution for the Dirac equation~(\ref{fieldDirac}), can be obtained straightforwardly.
Several relevant physical observables can then be found immediately,
such as the probability density, the transmission and reflection
coefficients between magnetic domains, and the density matrix, which
are all useful, for example, in graphene applications as those we
discuss below, where the Ritus method plays useful.

%With the ${\mathbb E}_p$ functions, we can straightforwardly construct the solutions to the Dirac equation, as we shall in the next section.
%
%\section{Solutions of the Dirac equation}
%
%Solutions to the Dirac equation in a background field~(\ref{fieldDirac}) are derived straightforwardly if we make the following ansatz
%\be
%\Psi = \mathbb {E}_p(z) u_{\bar {p}},\label{ansatz}
%\ee
%where $u_{\bar{p}}$ verifies the free Dirac equation~(\ref{free})
%\be
%(\gamma^\mu \Pi_\mu - m) \mathbb {E}_p(z) u_{\bar {p}}  = \mathbb {E}_p(z) (\gamma^\mu \bar {p}_\mu - m)  u_{\bar {p}}= 0. \ee
%with $\bar{p}_\mu=(p_0,0,\sqrt{k})$. Hence, the explicit solutions to the Dirac equation in an external uniform magnetic field are
%\bea
%\nn
%\psi_1 = \mathbb {E}_p u_{\bar{p}1}^{(A)} = \left(\begin{array}{c} E_{p,1}  \\ \frac {\sqrt {k}}{p_0 + m}E_{p,-1} \end{array}\right)
%\\
%\psi_2 = \mathbb {E}_p u_{\bar{p}2}^{(A)} = \left(\begin{array}{c}  - \frac {\sqrt {k}}{p_0 - m}E_{p,1}  \\ E_{p,-1} \end{array}\right)\label{solsdirac}
%\eea
%Needless to say, these expressions are no longer eigenfunctions of $\sigma_3/2$, which makes the notion of spin states cumbersome. To have a better taste of what these expressions mean, below we consider the case of a uniform magnetic field

\section{Examples}
\label{Examples}
\subsection{Uniform Magnetic Field}

\begin{figure}[t!] % fig1
\vspace{0.4cm}
{\centering
\resizebox*{0.45\textwidth}
{0.20\textheight}{\includegraphics{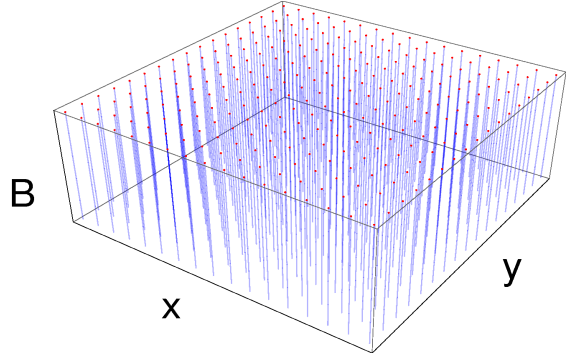}}
\par}
\caption{
The sketch of the field lines of a uniform magnetic field.
}
\label{fig1}
\end{figure}

\begin{figure}[t!] % fig2
\vspace{0.4cm}
{\centering
\resizebox*{0.45\textwidth}
{0.20\textheight}{\includegraphics{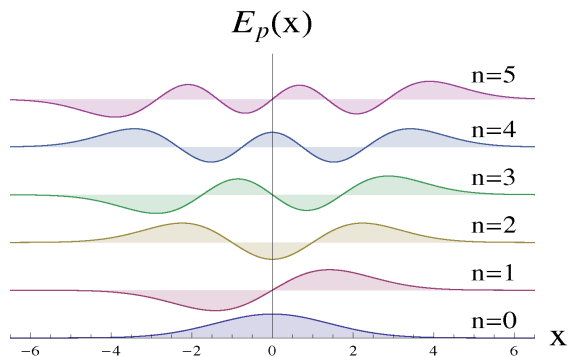}}
\par}
\caption{
The $E_p$ functions, Eq.(\ref{EpUnif}), in arbitrary units along the
dynamical direction for various Landau levels. The scale of the
graphic is set by $eB_0=1$ and $p_2=0$.
}
\label{fig2}
\end{figure}

To have an explicit taste of the power of Ritus method, let us first
consider the case of a uniform magnetic field.~\cite{Khalilov} This
corresponds to the choice $W(x)= B_0 x$ and the field lines are sketched in Fig.~\ref{fig1}. To simplify the calculations,
we rename the quantum number $k \to 2 \vert eB_0 \vert k$ in
Eq.~(\ref{eigenvalues}). In this case, Eq.~(\ref{pauli}) simplifies to
%%%%%%%%%%%%%%%%%%%%%%%%%%%%%%%%%%%%%%%%%%%%%%%%%%%%%%%%%%%%%%%%%%%%%%%%%%%%
%%%%%%%%%%%%%%%%%%%%%%%%%%%%%%%%%%%%%%%%%%%%%%%%%%%%%%%%%%%%%%%%%%%%%%%%%%%%%%

\be
\left[\partial_x^2 - (-p_2 + eB_0x)^2 + \sigma eB_0 \right]F_{k,p_2,\sigma}(x)=-2 \vert eB_0 \vert k F_{k,p_2,\sigma}(x)\;.
\ee
Let $\eta=\sqrt {2\vert eB_0 \vert}(x-\frac{p_2}{eB_0})$. Then the above
expression acquires the form
\be
\left[\frac{\partial^2}{\partial \eta^2} + k + \frac {\sigma}{2} \text{sgn} (eB_0) - \frac {\eta^2}{4}\right] F_{k,p_2,\sigma}(\eta)=0 \;,
\ee
that is, the equation for a quantum harmonic oscillator, with center of
oscillation in $x_0=p_2/(eB_0)$ and cyclotron frequency $w_c=
2eB_0$. Thus, the normalized functions $E_{p,\sigma}$ acquire the form
%\bea
%\nn
%E_{p,1} &=& \frac{(2\pi \vert e B \vert)^{1/4}}{(2 \pi)^{3/2}k!^{1/2}} e^{-ip_0 t + i p_2 y} D_k \left(\frac {\eta}{\sqrt {2}}\right) \;, \\
%\nn
%E_{p,-1} &=& \frac{(2\pi \vert e B \vert)^{1/4}}{(2 \pi)^{3/2}(k-1)!^{1/2}} e^{-ip_0 t + i p_2 y} D_{k-1} \left(\frac {\eta}{\sqrt {2}}\right)\;,
%\eea
\bea
\nn
E_{p,1} &=& \frac{(\pi \vert e B_0 \vert)^{1/4}}{2 \pi^{3/2} k!^{1/2}} e^{-ip_0 t + i p_2 y} D_k (\eta) \;, \\
E_{p,-1} &=& \frac{(\pi \vert e B_0 \vert)^{1/4}}{2 \pi^{3/2} (k-1)!^{1/2}} e^{-ip_0 t + i p_2 y} D_{k-1} ( \eta)\;,\label{EpUnif}
\eea
where 
\be
D_{n}(x)= 2^{-n/2}e^{-x^2 /4}H_n\left(x/\sqrt{2}\right)
\ee
is the
parabolic cylinder function of order 
\be
n=k +\frac{\sigma}{2}\text{sgn}(eB_0)- \frac{1}{2}\;,
\ee 
and $H_n(x)$ are the Hermite's polynomials. Expectedly, the uniform
magnetic field renders the $(n-1)$-th state with spin down with the
same energy of the $n$-th state with spin up.  In Fig.~\ref{fig2}, we depicted the functions $E_p\equiv E_{p,1}$  along the dynamical direction for various Landau levels. 
%%%%%%%%%%%%%%%%%%%%%%%%%%%%%%%%%%%%%%%%%%%%%%%%%%%%%%%%%%%%%%%%%%%%%%%%%%%%
%%%%%%%%%%%%%%%%%%%%%%%%%%%%%%%%%%%%%%%%%%%%%%%%%%%%%%%%%%%%%%%%%%%%%%%%%%%%%%

\begin{figure}[t!] % fig3
\vspace{0.4cm}
{\centering
\resizebox*{0.45\textwidth}
{0.20\textheight}{\includegraphics{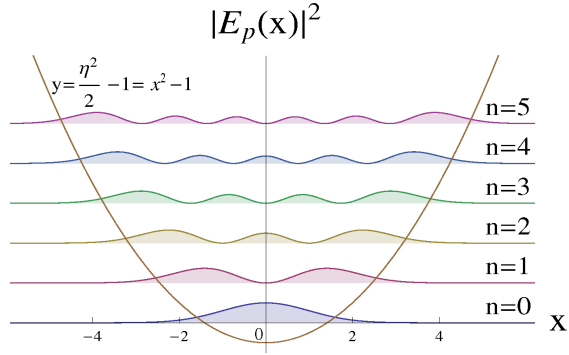}}
\par}
\caption{
The $|E_p|^2$ function from Eq.~(\ref{EpUnif}) in arbitrary units
along the dynamical direction for various Landau levels. The curve
$y=x^2-1$ corresponds to the potential~(\ref{suppot}) for this field
configuration. The scale of the graphic is set by $eB_0=1$ and
$p_2=0$.
}
\label{fig3}
\end{figure}

As we mentioned before, the Ritus eigenfunctions correspond to the asymptotic states of electrons in the presence of a magnetic field. Therefore, physical observables like probability densities are linear combinations of $|E_{p}|^2$. These functions have the profile shown  Fig.~\ref{fig3}. The solid curve enveloping these solutions corresponds to the potential
\be
y=\bar{W}^2(x)-\bar{W}'(x)\;,\label{suppot}
\ee
where $\bar{W}=eW-p_2$ is
referred to as the superpotential in the SUSY-QM literature.~\cite{susyqm}  

%%%%%%%%%%%%%%%%%%%%%%%%%%%%%%%%%%%%%%%%%%%%%%%%%%%%%%%%%%%%%%%%%%%%%%%%%%%%
%%%%%%%%%%%%%%%%%%%%%%%%%%%%%%%%%%%%%%%%%%%%%%%%%%%%%%%%%%%%%%%%%%%%%%%%%%%%%%

\subsection{Exponential Magnetic Field}

There are many problems relating electrons in non-uniform
magnetic fields of relevance in graphene.  In particular, it has been established the possibility to confine quasiparticles in magnetic barriers.~\cite{gaby} This could be feasible creating 
spatially inhomogeneous, but constant in time, magnetic fields depositing ferromagnetic layers over the substrate of a graphene sample layer.~\cite{Reijniers}
The physical properties of graphene make it a promising novel material to control the transport properties in nanodevices. It has been considered to 
be used in electronics and spintronics applications, like in single-electron transistors,~\cite{Ponomarenko}  in the so called magnetic edge states,~\cite{Park} which may play an important role in the spin-polarized currents along magnetic domains, and in quantum dots and antidots magnetically confined.
Moreover, the quantum Hall effect in graphene has been observed at room temperature,~\cite{Novoselov2}  evidence which confirms the great potential of graphene as the material to be used in carbon-based electronic devices.

\begin{figure}[t!] % fig4
\vspace{0.4cm}
{\centering
\resizebox*{0.45\textwidth}
{0.20\textheight}{\includegraphics{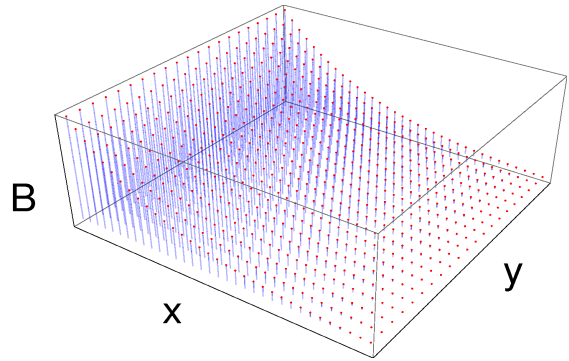}}
\par}
\caption{
The sketch of the field lines of an exponentially decaying static
magnetic field.
}
\label{fig4}
\end{figure}

\begin{figure}[t!] % fig5
\vspace{0.4cm}
{\centering
\resizebox*{0.45\textwidth}
{0.20\textheight}{\includegraphics{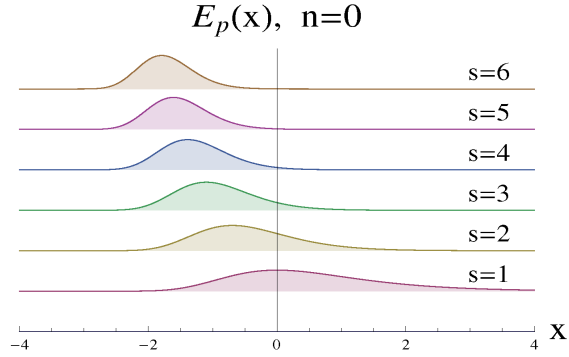}}
\par}
\caption{
The $E_p$ functions from Eq.~(\ref{EpExp}) in arbitrary units along
the dynamical direction for various values of $s$ at fixed $n=0$. The
scale of the graphic is set by $eB_0=\alpha=1$.
}
\label{fig5}
\end{figure}

\begin{figure}[t!] % fig6
\vspace{0.4cm}
{\centering
\resizebox*{0.45\textwidth}
{0.20\textheight}{\includegraphics{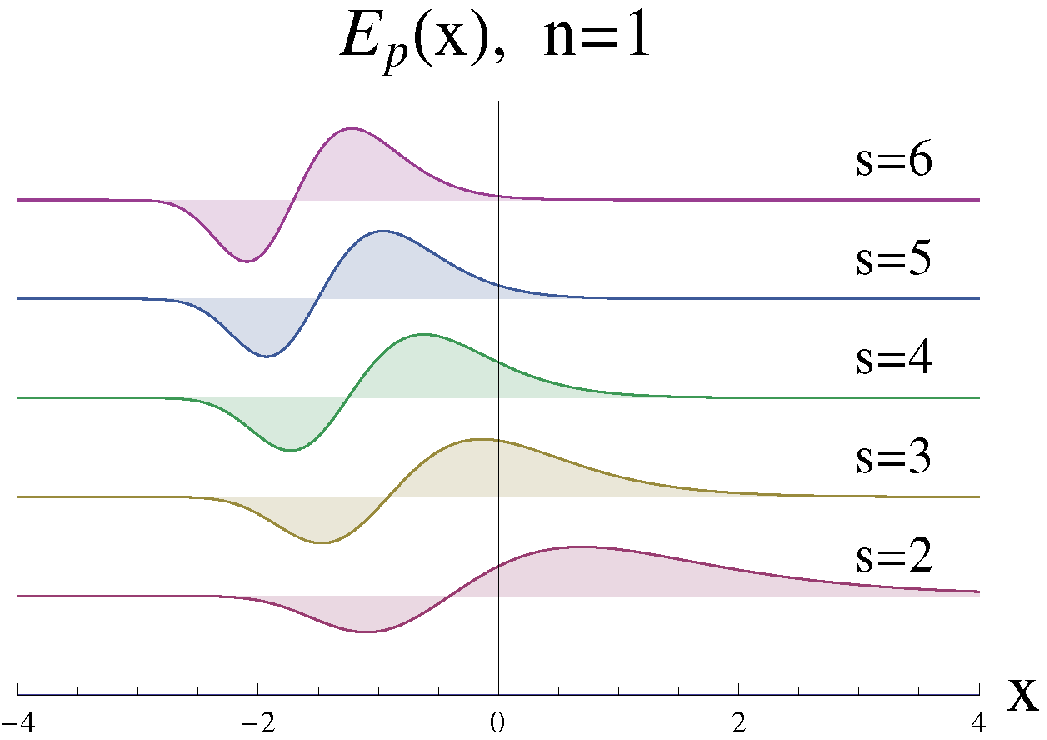}}
\par}
\caption{
The $E_p$ functions from Eq.~(\ref{EpExp}) in arbitrary units along
the dynamical direction for various values of $s$ at fixed $n=1$. The
scale of the graphic is set by $eB_0=\alpha=1$.
}
\label{fig6}
\end{figure}

\begin{figure}[t!] % fig7
\vspace{0.4cm}
{\centering
\resizebox*{0.45\textwidth}
{0.20\textheight}{\includegraphics{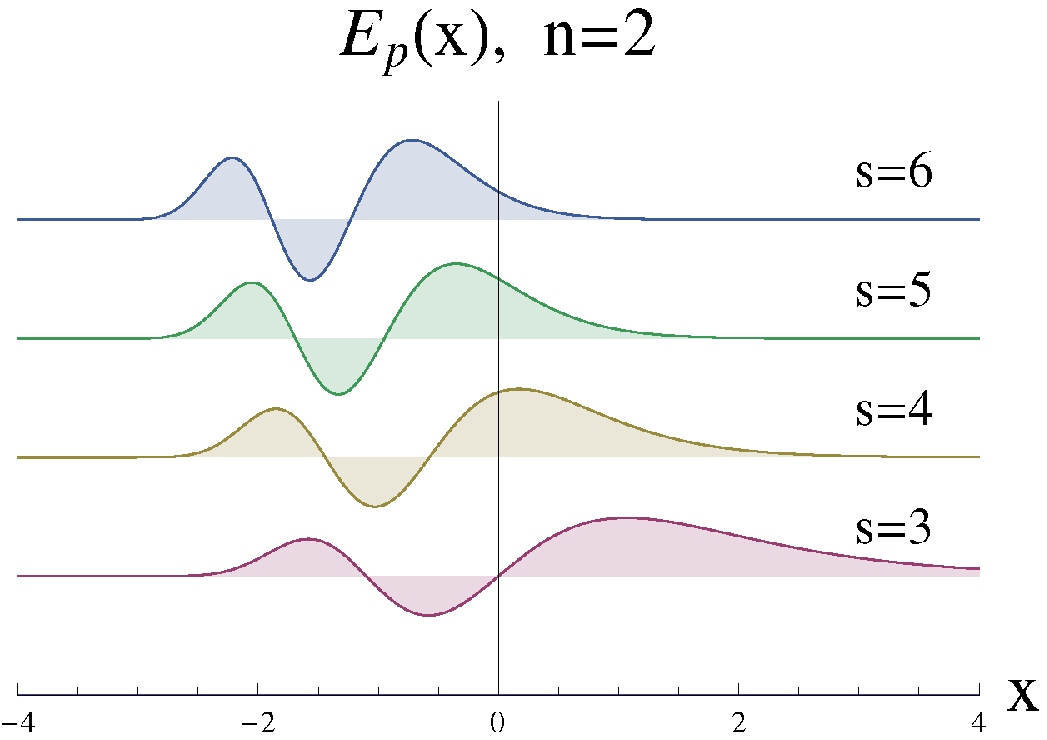}}
\par}
\caption{
The $E_p$ functions from Eq.~(\ref{EpExp}) in arbitrary units along
the dynamical direction for various values of $s$ at fixed $n=2$. The
scale of the graphic is set by $eB_0=\alpha=1$.
}
\label{fig7}
\end{figure}

\begin{figure}[t!] % fig8
\vspace{0.4cm}
{\centering
\resizebox*{0.45\textwidth}
{0.20\textheight}{\includegraphics{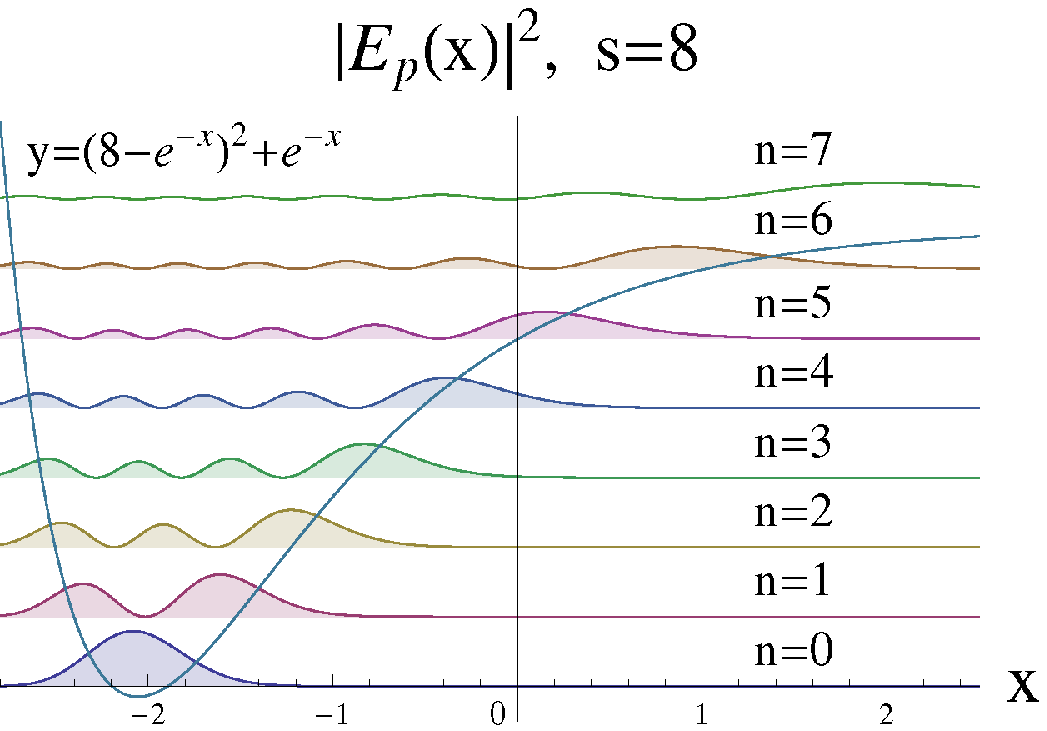}}
\par}
\caption{
The $|E_p|^2$ functions from Eq.~(\ref{EpExp}) in arbitrary units
along the dynamical direction for various values of $n$ at fixed
$s=8$. The curve $y=(8-e^{-x})^2+e^{-x}$ corresponds to the
potential~(\ref{suppot}) for this field configuration. The scale of
the graphic is set by $eB_0=\alpha=1$.
}
\label{fig8}
\end{figure}

Here we study the electron propagator in a background static magnetic field which has an exponentially decaying spatial profile along one direction, described through the function $W(x) = -B_0\exp\{-\alpha x\}$. The field lines of such magnetic configuration are sketched in Fig.~\ref{fig4}.
In this case, Eq.~(\ref{pauli}) simplifies to
%%%%%%%%%%%%%%%%%%%%%%%%%%%%%%%%%%%%%%%%%%%%%%%%%%%%%%%%%%%%%%%%%%%%%%%%%%%%
%%%%%%%%%%%%%%%%%%%%%%%%%%%%%%%%%%%%%%%%%%%%%%%%%%%%%%%%%%%%%%%%%%%%%%%%%%%%%%
\be
\left[  \partial_x^2 
      - (-p_2 - eB_0 \exp\{-\alpha x\} )^2 
      + \sigma \alpha eB_0 \exp \{-\alpha x\}
\right]F_{k,p_2,\sigma}(x)=- k F_{k,p_2,\sigma}(x)\;.
\ee
Let $\varrho = (2eB_0/\alpha)\exp \{ -\alpha x\}$ and
$s=-p_2/\alpha$, then, the above expression is equivalent to
\be
\left[   \varrho^2 \frac{\partial^2}{\partial \varrho^2} 
       + \varrho \frac{\partial}{\partial \varrho}
       - \left(s-\frac{1}{2}\varrho\right)^2 
       +  \frac{\sigma}{2}\varrho 
       + \frac{k}{\alpha^2} 
\right] F_{k,p_2,\sigma}(\varrho)=0 \;.
\ee
This equation has the normalized solutions $E_{p,\sigma}$ given as
\bea
\nn
E_{p,1} &=& \frac{1}{2\pi}\left( \frac{2 \alpha n! (s-n)}{\Gamma (2s-n+1)} \right)^{1/2} e^{-ip_0 t + i p_2 y} e^{-\varrho/2} \varrho^{(s-n)}L_{n}^{2(s-n)} (\varrho) , \\
E_{p,-1} &=& \frac{1}{2\pi}\left( \frac{2 \alpha (n-1)! (s-n)}{\Gamma (2s-n)} \right)^{1/2} e^{-ip_0 t + i p_2 y} e^{-\varrho/2} \varrho^{(s-n)}L_{n-1}^{2(s-n)} (\varrho),\label{EpExp}
\eea
where $L_a^b(x)$ are the associate Laguerre polynomials with
\be
n=s-\sqrt{-\frac{k}{\alpha^2} + s^2}.
\ee
 The quantum number $n$ is the
principal quantum number, whereas $s$ a center of oscillation weighted
by the damping factor $\alpha$.  The solutions $E_p$ from Eq.~(\ref{EpExp}) are sketched in Figs.~\ref{fig5}-\ref{fig7} for $n=0,1$ and 2 for various values of $s$, whereas in Fig.~\ref{fig8} we show $|E_p|^2$ for various values of $n$ at fixed $s=8$.
Notice that in this case the potential~(\ref{suppot}) also envelops the squares of the solutions.

%%%%%%%%%%%%%%%%%%%%%%%%%%%%%%%%%%%%%%%%%%%%%%%%%%%%%%%%%%%%%%%%%%%%%%%%%%%%
%%%%%%%%%%%%%%%%%%%%%%%%%%%%%%%%%%%%%%%%%%%%%%%%%%%%%%%%%%%%%%%%%%%%%%%%%%%%%%

%%%%%%%%%%%%%%%%%%%%%%%%%%%%%%%%%%%%%%%%%%%%%%%%%%%%%%%%%%%%%%%%%%%%%%%%%%%%
%%%%%%%%%%%%%%%%%%%%%%%%%%%%%%%%%%%%%%%%%%%%%%%%%%%%%%%%%%%%%%%%%%%%%%%%%%%%%%

%%%%%%%%%%%%%%%%%%%%%%%%%%%%%%%%%%%%%%%%%%%%%%%%%%%%%%%%%%%%%%%%%%%%%%%%%%%%
%%%%%%%%%%%%%%%%%%%%%%%%%%%%%%%%%%%%%%%%%%%%%%%%%%%%%%%%%%%%%%%%%%%%%%%%%%%%%%

%%%%%%%%%%%%%%%%%%%%%%%%%%%%%%%%%%%%%%%%%%%%%%%%%%%%%%%%%%%%%%%%%%%%%%%%%%%%
%%%%%%%%%%%%%%%%%%%%%%%%%%%%%%%%%%%%%%%%%%%%%%%%%%%%%%%%%%%%%%%%%%%%%%%%%%%%%%

This concludes our presentation of the Ritus method in magnetic
fields. In the next section we consider the Schwinger model
for electric fields within this approach.

\section{Propagator in Electric Fields}
\label{Propagator_Electric_Fields}

The Dirac equation in the massive Schwinger
model~\cite{Schwinger} takes the form shown in Eq.~(\ref{fieldDirac}), but only two $\gamma^\mu$-matrices are involved. In this section, we
work in a (1+1)-dimensional Minkowski space with metric
$g^{\mu\nu}=g_{\mu\nu}=\text{diag}(1,-1)$, again in natural units. We
choose the $\gamma^\mu$-matrices as 
\bea
\gamma_0=\left(\begin{tabular}{c c} 0 & 1\\ 1 & 0
                 \end{tabular}
           \right) \;, \qquad
  \gamma_1=\left(\begin{tabular}{c c}
                  0 & -1\\
                  1 & \phantom{-}0
                 \end{tabular}
           \right)\; .\eea
Notice that           
           \bea
                 \gamma_5=\gamma_0\gamma_1=\left(\begin{tabular}{c c}
                  1 & \phantom{-}0\\
                  0 & -1
                 \end{tabular}
           \right)\equiv\sigma_{01} \;.
\eea
We introduce a constant uniform electric field in the gauge
\bea
    A_\mu=(-E_0 x, 0)\;,
\eea
and then, since the only non-vanishing components of the field
strength tensor are $F_{01}=-F_{10}=E_0$, we have that \bea
(\gamma\cdot\Pi)^2{\mathbb E}_p&=& \left[\Pi^2
  -ie\gamma_5E_0\right]\mathbb{E}_p \ = \ p^2\mathbb{E}_p\;.  \eea

We look for solutions of the form
\bea
\mathbb{E}_p&=&E_{p,\sigma}(t,x)\omega_\sigma
\ = \ e^{-ip_0t}\ E_{p,\sigma}(x)\omega_\sigma \;,
\eea
where $\omega_\sigma$ is the matrix of eigenvectors of $\gamma_5$ with
eigenvalues $\sigma=\pm1$, respectively. Explicitly, the Ritus eigenfunctions for
this problem are of the form
\bea
 \mathbb{E}_p=e^{-ip_0t}\left(
                 \begin{tabular}{c c}
                 $E_{p, 1}(x)$ & 0 \\
                  0       & $E_{p,-1}(x)$
                 \end{tabular}
                \right)\;,
\eea
where the $E_{p,\sigma}$ functions verify
\bea
\left[-(\partial_t+ieE_0 x)^2+\partial_x^2
               -ie\sigma E_0\right]E_{p,\sigma}= p^2E_{p,\sigma}\;.
\eea
%\bea
%\left[\partial_x^2+(eE_0)^2\left(x+\frac{p_0}{eE_0}\right)^2\right]E_{p,\sigma}= \tilde{p}^2E_p^\sigma
%\eea
%where $\tilde{p}^2=2ieE_0 (p^2-\sigma/2)$. 
Let us make the replacement $p^2\to 2ieE_0 p^2$ as in the case of a
uniform magnetic field. Then making the change of variables
$E_0\rightarrow i{\cal E}_0$ and $\rho=\sqrt{2e{\cal
    E}_0}\left(x+ip_0/(e{\cal E}_0)\right)$, we get
\bea
\left[\frac{\partial^2}{\partial\rho^2}+p^2+\frac{\sigma}{2}-\frac{\rho^2}{4}\right]E_{p,\sigma}(\rho)=0 \; , 
\eea 
which is again a displaced harmonic oscillator with center of
oscillation in $x_0=ip_0/(e{\cal E}_0)$ and frequency
$\omega_c=2e{\cal E}_0$. Thus the solutions are again parabolic
cylinder functions
%{\color{red}\bea
%E_{p,\sigma}(\rho)=\left(\frac{4e{\cal E}_0}{\pi}\right)^{\frac{1}{2}}
%      \frac{1}{\sqrt{2^nn!}}H_n(\rho)e^{-\frac{\rho^2}{2}}
%       \equiv D_n(\rho)
%\eea}
\bea
E_{p,1}(\rho)&=& \frac {(2e {\cal E}_0)^{1/4}
}{(2\pi)^{3/2} (n+1)!^{1/2}} D_{n+1}(\rho)\;, \nn\\
E_{p,-1}(\rho)&=& \frac {(2e {\cal E}_0)^{1/4}
}{(2\pi)^{3/2} n!^{1/2}} D_{n}(\rho)\;, 
\eea
of order
\be
n=p^2+\frac{\sigma}{2}-\frac{1}{2}.\label{orderel}
\ee
Contrary to the case of the uniform magnetic field, the arguments of
the parabolic cylinder functions as well as the
order, are complex numbers. Nevertheless, they serve
as well to diagonalize the propagator as

\bea
S(x)=\int dp \ \mathbb{E}_p(x)\frac{1}{\gamma\cdot \overline{p}-m}
                \overline{\mathbb{E}}_p(x) \;,
\eea
which is found through the relation~(\ref{p_barra}) with $\overline{p}_\mu=(p,0)$, where $p$ is defined in Eq.~(\ref{orderel}).
%Obviously, the massless version of this propagator corresponds to the
% propagator of the Schwinger Model.

This illustrates the usefulness of the Ritus method. Below we present
a discussion and the conclusions.

\section{Discussion and conclusions}
\label{Conclusions}

In this paper we have reviewed the massive electron propagator in
lower than (3+1) space-time dimensions. We have
obtained the
%solutions $\psi(x)$ to the free Dirac equation, which consist of only
%a particle solution with spin up, and an antiparticle with spin
%down. Solutions for the second irreducible representation can be
%straightforwardly derived from these through the replacement
%$\psi'\to\sigma_1\psi$ and comprise a particle solution with spin
%down and an antiparticle solution with spin up. Although these two
%irreducible set of spinors can be merged into a reducible 4-component
%spinor, we do not pursue this line and prefer to leave as an
%excersice to the reader the proper generalization. The
free electron propagator through the Green's function method in
arbitrary dimensions and has, as we know, a diagonal form in momentum
space, namely $S_F(p)=1/(\gamma\cdot p-m)$.  This is obviously the case
because the free Dirac operator commutes with the momentum operator
and thus one can take simply a Fourier transform, that is to say, we
can use the eigenfunctions of the momentum operator to diagonalize the
Green's function.

We have then considered the electron propagator in
(2+1)-dimensions in an external magnetic field pointing
perpendicularly to the plane of motion of the electron and whose
spatial shape we considered described by some arbitrary function
$W(x)$. This, to the best of our knowledge, has been presented for the
first time in literature. We selected to work within the Ritus
method~\cite{Ritus} framework, and construct explicitly the
eigenfunctions ${\mathbb E}_p$ of the $\gp^2$ operator in the general
case. The basic idea behind this method is the following: since the
Dirac operator $\gp$ does not commute with the momentum operator, we
cannot simply take the Fourier transform of the Green's function and
have a diagonal propagator in momentum space. However since $\gp$
commutes with $i\partial_t$, $-i\partial_y$ and, more importantly,
with $\gp^2$, we can use their common
eigenfunctions, ${\mathbb E}_p$, to diagonalize the propagator, which
turns out to have the form of a free propagator with momentum
depending upon the dynamical quantum
numbers. We have specialized our findings to the case of a uniform
magnetic field where the ${\mathbb E}_p$ functions are described in
terms of parabolic cylinder functions. As a second
example, we considered an exponentially damped static magnetic field. Here
the Ritus eigenfunctions are written in terms of associate Legendre
functions. On both these cases, the massless versions of the
propagators are of direct relevance in graphene.

%Solutions to the Dirac equation have also been derived in sect. IV.
%Their form is obtained from the ansatz in Eq.~(\ref{ansatz}) and
%hence they are labeled by the quantum numbers of the ${\mathbb E}_p$
%functions presented in Eq.~(\ref{eigenvalues}). In order to taste the
%power and usefulness of the method, Other configurations of external
%fields can be considered in this formalism, so long as one is able to
%solve Eq.~(\ref{pauli}). The price to pay is a more complicated form
%of the ${\mathbb E}_p$ functions, although the propagator itself
%would continue to be diagonal, with the form of a free propagator.

We also considered the propagator in (1+1)-dimensions
in a uniform electric field. Our findings in this case are similar to
the magnetic field case in the plane, except that the eigenfunctions
have complex arguments and order. The important lesson here is that
the method works in arbitrary dimensions and different field
configurations. In each case, the propagator continues to have a
diagonal form, although the ${\mathbb E}_p$ functions change from one
configuration to the other.

In conclusion, the Ritus method offers an alternative to study the
electron propagator in the presence of external fields. Contrary to
the ``proper time'' method~\cite{Fock,Schwinger}, the
propagator here acquires its free form,  and the effects
of the field are reflected in the eigenfunctions of the operator
$\gp^2$, namely ${\mathbb E}_p$. 
The method works as well in the case of scalar~\cite{scalar} and vector charged particles.~\cite{vector} The idea behind it is familiar in
quantum mechanics, where the diagonalization of a given operator is
carried out with the aid of eigenfunctions of a
second commuting operator. That makes Ritus method suitable to study
the propagator for more complex configurations of external fields.

\section*{Project}

For the interested reader, we propose the following project: Work out the derivation of the propagator in (2+1)-dimensions in a magnetic field making use of the second irreducible representation of the Dirac matrices, Eq.(\ref{segunda}), and show that in this case, the ${\mathbb E}_p$ functions have the form
\be
{\mathbb E}_p=\left(\begin{array}{cc} E_{p,-1}(z) & 0 \\ 0 & E_{p,1}(z)\end{array} \right)\;.
\ee

\begin{acknowledgments}
The authors are indebted to Alejandro Ayala and Adnan Bashir for valuable discussions and careful
reading of the manuscript. AR acknowledges support from SNI and
CONACyT grants under project 82230.
\end{acknowledgments}


\begin{thebibliography}{99}
\bibitem{fn1} We use a prescription for which $\gamma\cdot
  A=\gamma^\mu A_\mu=\gamma_\mu A^\mu$ for any vector $A^\mu$ in $d$-dimensions.
\end{thebibliography}
\end{document}